\documentclass[amsmath,amssymb,aps,pra,reprint,showpacs,floats,superscriptaddress,longbibliography]{revtex4-2}
    \usepackage[utf8]{inputenc}
    \usepackage[table,xcdraw]{xcolor}
    \usepackage{graphicx}
    \usepackage{amsmath, amsthm, amssymb, amsfonts}
    \usepackage{mathtools}
    \usepackage[colorlinks=true,citecolor=blue,linkcolor=blue,urlcolor=blue]{hyperref}
    \usepackage{footnote}
    \usepackage{xifthen}    
    \usepackage{capt-of}
    \usepackage{subfigure}
    \usepackage{physics}
    \usepackage{adjustbox}
    \usepackage{lipsum}
    \usepackage{quantikz}
    \usepackage{tabularx}
    \usepackage{siunitx}
    \usepackage{booktabs}
    \pdfoutput=1 
    \definecolor{mplblu}{HTML}{1F77B4}
    \definecolor{mplorg}{HTML}{FF7F0E}
    \definecolor{mplgrn}{HTML}{2CA02C}
    \definecolor{mplred}{HTML}{D62728}
    \definecolor{mplpur}{HTML}{9467BD}

    \newcommand{\ntscite}[1][]{{\tiny \textcolor{blue}{\ifthenelse{\equal{#1}{}}{[CITE]}{[CITE: #1]}}}}
    \newcommand{\ntsref}[1][]{{\tiny \textcolor{blue}{\ifthenelse{\equal{#1}{}}{[REF]}{[REF: #1]}}}}

    \newcommand{\cbqty}[1]{\left\{#1\right\}}

    \newcommand{\sn}[1]{\mathcal{#1}}
    \newcommand{\opr}[1]{#1}

    \newcommand{\HH}{\opr{H}}
    \newcommand{\VV}{\opr{V}}

    \newcommand{\bbi}{\opr{\mathbb{I}}}

    \newcommand{\pA}[1][]{\ifthenelse{\equal{#1}{}}{\mathcal{A}}{\mathcal{A}^{(#1)}}}
    \newcommand{\pB}[1][]{\ifthenelse{\equal{#1}{}}{\mathcal{B}}{\mathcal{B}^{(#1)}}}
    \newcommand{\pC}[1][]{\ifthenelse{\equal{#1}{}}{\mathcal{C}}{\mathcal{C}^{(#1)}}}

    \newcommand{\ssblabel}{\sn{B}}
    \newcommand{\ssclabel}{\sn{S}}

    \newcommand{\bbiB}{\bbi^\ssblabel}
    \newcommand{\bbiS}{\bbi^\ssclabel}

    \newcommand{\NE}{N_{\sn{E}}}

    \newcommand{\HHS}{\HH^\sn{S}}
    \newcommand{\HHE}{\HH^\sn{E}}

    \newcommand{\ncl}{\mathcal{N}}

    \newcommand{\TR}[1]{\mathrm{tr}\left\{#1\right\}}
    \newcommand{\IM}[1]{\mathfrak{I}\left\{#1\right\}}
    \newcommand{\RE}[1]{\mathfrak{R}\left\{#1\right\}}

\begin{document}
%
\title{From compatibility of measurements to exploring Quantum Darwinism on NISQ}
\author{Emery~Doucet}
\email{emery.doucet001@umb.edu}
\affiliation{Department of Physics, University of Massachusetts, Boston, Massachusetts 02125, USA}
\affiliation{Department of Physics, University of Maryland, Baltimore County, Baltimore, MD 21250, USA}
\affiliation{Quantum Science Institute, University of Maryland, Baltimore County, Baltimore, MD 21250, USA}
\author{Sebastian~Deffner}
\email{deffner@umbc.edu}
\affiliation{Department of Physics, University of Maryland, Baltimore County, Baltimore, MD 21250, USA}
\affiliation{Quantum Science Institute, University of Maryland, Baltimore County, Baltimore, MD 21250, USA}
\affiliation{National Quantum Laboratory, College Park, MD 20740, USA}
\date{\today}
\begin{abstract}
Quantum Darwinism explains how tenets of classical reality, such as objectivity and repeatability, emerge within a quantum universe. As a mathematical framework, Quantum Darwinism also provides guiding principles that determine what physical models support emergent classical behavior, what specific observables obey classical laws, and much more. For instance, in a recent work we elucidated that the limit under which  Kirkwood-Dirac quasiprobability distributions become effectively classical coincides with the regime where the underlying physical model obeys the rules of Quantum Darwinism. In the present work, we study the breaking of Quantum Darwinism in a specific model and how that translates to non-classical measurement statistics. Interestingly, this provides effective tools for benchmarking the genuine quantum characteristics of NISQ hardware, which we demonstrate with IonQ's trapped-ion and IBM's superconducting quantum computing platforms.
\end{abstract}
\maketitle
%

%
%
Quantum physics provides a fundamental description of the rules that determine how information can be processed and communicated in our Universe \cite{Campbell2025QST}.
As a result of the large-scale investment into quantum information processing technologies in recent years, it is now possible to harness these foundational processes for computational tasks.
One of the primary motivations of quantum computing research is that programmable quantum computers enable the study of all different types of quantum systems in their natural quantum habitat, through quantum simulation \cite{Georgescu2014,Buluta2009,Daley2022}.

In the present work, we consider quantum simulation as a tool to study how classical behavior emerges from quantum systems.
The framework which describes this process in general is called Quantum Darwinism \cite{Zurek03, Zurek09, BlumeKohout05, BlumeKohout06, Ollivier04, Ollivier05, Riedel2010PRL, Riedel2011NJP, Brandao2015NC, Knott2018PRL, Milazzo2019PRA, Colafranceschi2020JPA, Touil22b, Girolami2022PRL, Zurek22, QDAndFriends, Zurek2000, Zurek22, Kiely2025, Touil2025}, and it provides a concrete explanation of precisely how features of the classical world and in particular classical measurements (repeatability, objectivity, etc.) manifest as emergent behaviors in quantum systems interacting with an environment.
The key insight is that ``real'' measurements are always indirect -- observers do not probe a target system directly, instead that system interacts with its environment and some fragment of that environment is captured to provide information on the system state. 

Despite its theoretical success and conceptual importance, experimental exploration of the framework of Quantum Darwninism is notoriously difficult \cite{Ciampini2018PRA,Unden2019PRL,Chen2019SB,Chisholm2022QST}. A particularly promising route relies on verifying the special form of the emergent pointer states \cite{Zhu2025SA}, which we used to uniquely specify the physical models that can support the emergence of classical behavior \cite{Doucet24}. Exploiting this insight, we recently developed a deep connection between Quantum Darwinism and Kirkwood-Dirac (KD) quasiprobability distributions \cite{Kirkwood1933, Dirac1945} for multi-party measurements in Ref.~\cite{Doucet25}.
The KD quasiprobability distribution is a generalization of the classical joint probability distributions taking into account the possibility that quantum measurements may be incompatible.
In Ref.~\cite{Doucet25}, we showed that the set of models which exhibit Quantum Darwinism is precisely the same as the set of models where the KD quasiprobability distributions of all possible sets of disjoint measurements reduce to their classical probability counterparts.
That is, one may equivalently interpret Quantum Darwinism through the lens of the objectivity of specific pointer observables \cite{Duruisseau23,Touil24,Doucet24}, or of the compatibility of all sets of measurements \cite{Doucet25}. 

This second perspective is interesting because it can be quantified mathematically as well as experimentally on NISQ devices; there exist a number of different measures of the non-classicality $\ncl$ of a given KD quasiprobability distribution \cite{ArvidssonShukur21,GonzalezAlonso19,He24}.
When they are non-zero, quantum correlations are present which can be diagnosed using the corresponding measurements, meaning there exists some protocol using these measurements which can witness the ``quantum-ness'' of the system.
When the measures vanish, the given measurements cannot witness non-classical behavior. 
Hence the connection to Quantum Darwinism and emergent classicality -- when no possible set of measurements can witness non-classical behavior, the system is effectively classical.
If even a single measurement setting can be exhibited which witnesses non-classical behavior, the model is not Darwinistic. 

Generally, we expect in the non-Darwinistic case that ``most'' measurements should exhibit non-zero $\ncl$ and so witness non-classicality.
Intuitively this is because the absence of Quantum Darwinism (equivalently, the presence of KD non-classicality) is typically associated with entanglement growth coupled with some kind of scrambling dynamics \cite{Duruisseau23,Doucet24,Touil20,Swingle18,Touil22,Mohseninia2019,GonzalezAlonso19}. 
Over time, this should spread the support of measurement operators acting on a fragment of the environment over the full environment in a manner which is difficult to control and hence it is difficult to ensure that initially commuting measurements remain commuting at all times.
Of course the existence of symmetries can complicate this simple analysis, however with ``sufficiently random'' choices of measurement bases the conclusion holds.

In this work, we make use of this connection between Quantum Darwinism and non-classicality of KD quasiprobability distributions in order to probe the quantum-to-classical transition in a simple system experimentally using gate-based NISQ devices developed by IBM and IonQ. 
We construct quantum circuits to simulate the dynamics of a simple quantum system having one parameter which can be swept to control if the simulated system exhibits Quantum Darwinism or not, and if not to what degree.
For certain specific measurement settings, we choose a small set of parameters which are known to exhibit (non-)classical correlations and test each on both simulators and real quantum hardware by measuring the amount of non-classicality.
This serves as a useful and practical benchmarking tool for these devices, and in the future can be extended to larger and more complicated systems to provide new insights. 


%
%
\section{Background}
\label{sec:Background}

One of the defining features of quantum physics is that quantum measurements are not necessarily compatible with one another in the same way that classical measurements are.
This incompatibility is a consequence of quantum correlations that manifest in a number of ways \cite{Lostaglio2023,Arvidsson-Shukur2024NJP} and are deeply related to a number of uniquely quantum phenomena such, for instance scrambling of quantum information \cite{GonzalezAlonso19}, quantum thermodynamics \cite{YungerHalpern2017}, quantum speed limits \cite{Budiyono2025}, and more.

For our present purposes, the primary consequence of measurement incompatibility is the implication that it becomes impossible to write down a joint probability distribution describing a pair of incompatible measurements \cite{BreuerPetruccione}.
Assuming two joint measurements are performed with outcomes $\cbqty{A_i}$ and $\cbqty{B_j}$ respectively, the natural choice is
\begin{align}
    p_{ij} = \TR{B_j A_i \rho A_i B_j}\,.
\end{align}
However, this describes a two-point measurement protocol which presupposes an order of ``measurement A, followed by measurement B''.
This can cause issues if the measurements fail to commute, and leads to a joint probability distribution which fails to reproduce the correct marginal distributions \cite{BreuerPetruccione}. 

Instead, in the quantum world it is more natural to work with complex-valued quasiprobabilities such as the Kirkwood-Dirac quasiprobability distribution,
\begin{align}
    q_{ij} = \TR{B_j A_i \rho}\,,
    \label{eqn:KDQ}
\end{align}
which reproduces the correct marginal distributions \cite{Lostaglio2023,BreuerPetruccione,Spriet2025,Arvidsson-Shukur2024NJP,Budiyono2024}. The KD distribution is widely applicable throughout quantum mechanics, with applications in virtually all areas of quantum physics, cf. recent review articles \cite{Arvidsson-Shukur2024NJP,Gherardini2024PRXQ}.

The KD quasiprobability distribution can be written as the sum of the two-point measurement probability and a pair of quantum modification terms \cite{Johansen07},
\begin{align}
    q_{ij} = 
        p_{ij} 
        + \frac{1}{2}\TR{(\rho - \rho')B_j} 
        + \frac{i}{2}\TR{(\rho - \rho')B_j^{\pi/2}}
    ,
    \label{eqn:KDQWithModificationTerms}
\end{align}
where $\rho'$ is the state after a non-selective measurement of $A_i$ and its complement,
\begin{align}
    \rho' 
        &= A_i\rho A_i + \pqty{\bbi - A_i}\rho\pqty{\bbi - A_i}
    ,
\end{align}
and $B_j^{\pi/2}$ is the projector $B_j$ with phases adjusted in the $A_i$ basis,
\begin{align}
    B_i^{\pi/2} 
        &= e^{i \pi A_i / 2} B_j e^{-i \pi A_i /2}
    .
\end{align}
The quantum modification terms are zero if and only if the two measurements commute \cite{Johansen07}, implying that the KD quasiprobability distribution reduces to the classical two-point measurement joint probability distribution. 
If they are non-zero, quantum correlations are present and at least some of the individual quasiprobabilities $q_{ij}$ will either be negative or have non-zero imaginary components.

In Ref.~\cite{Doucet25} we showed that the limit in which the KD quasiprobability distribution becomes a proper probability distribution is governed by Quantum Darwinism.
Since the quantum modification terms ``witness'' quantum correlations, they provide an experimentally accessible tool for the diagnostics of the quantum-to-classical transition: if it can be shown that the modification terms are always identically zero for any collection of disjoint measurements on a composite system with many degrees of freedom, then that system is effectively classical.
%
%
\subsection{Measures of Non-classicality}

A number of methods to quantify the degree to which a given KD quasiprobability distribution exhibits non-classicality have been proposed \cite{GonzalezAlonso19,ArvidssonShukur21,He24}.
For example, one could use the norm of the quantum modification terms \cite{He24},
\begin{align}
\ncl_{H} = \frac{1}{2}\sum_{ij}\left( \left|\TR{(\rho - \rho')B_j}\right| + \left|\TR{(\rho - \rho')B_j^{\pi/2}}\right| \right),
\end{align}
as a measure of non-classicality, directly measuring the difference between the KD distribution and the associated classical distribution.
Unfortunately, $\ncl_{H}$ cannot be computed from only knowledge of the quasiprobabilities $q_{ij}$ alone.
Instead, it requires knowledge of the quasiprobabilities as well as the joint probabilities $p_{ij}$ corresponding to the two-point measurement protocol.
This is problematic from the point of view of experimentally quantifying the non-classicality, as at least two distinct measurement protocols must be tested and compared.

Instead, we turn to measures based on negativity or non-reality of the quasiprobabilities themselves \cite{GonzalezAlonso19,ArvidssonShukur21}.
Specifically, we consider the measure defined in Ref.~\cite{ArvidssonShukur21}, $\ncl_{\rm AS} \equiv \ncl_{\rm AS}^{\mathfrak{R}} + \ncl_{\rm AS}^{\mathfrak{I}}$ with, 
\begin{subequations}
\begin{align}
   \ncl_{\rm AS}^{\mathfrak{R}} &= \sum_{ij} \left|\RE{q_{ij}}\right| - 1
    ,
    \\
    \ncl_{\rm AS}^{\mathfrak{I}} &= \sum_{ij} \left|\IM{q_{ij}}\right|
    .
\end{align}
    \label{eqn:NclAS}
\end{subequations}

An alternative measure of non-classicality that could be useful in some contexts is obtained by taking the largest imaginary part or negative real part of any single quasiprobability,
\begin{subequations}
\begin{align}
    \ncl_{\infty}^{\mathfrak{R}} &= \max\big(0, -\min_{ij}\RE{q_{ij}}\big)
    ,
    \\
    \ncl_{\infty}^{\mathfrak{I}} &= \max_{ij}|\IM{q_{ij}}|
    ,
\end{align}
\end{subequations}
and combining them as $\mathcal{N}_\infty \equiv \ncl_{\infty}^{\mathfrak{R}} + \ncl_{\infty}^{\mathfrak{I}}$.
For the simple measurements we consider in the remainder of this work, in fact $\mathcal{N}_\infty$ is just a multiple of $\mathcal{N}_{\rm AS}$ and so there is no meaningful difference.
Nonetheless, from an experimental perspective we expect that this measure should be easier to perform error analysis for, and if the ``most non-classical'' quasiprobability is known before hand allows the non-classicality to be estimated by measuring only one quasiprobability. 

%
%
\subsection{Measuring quasiprobabilities}

A number of approaches exist or can be adapted to the measurement of KD quasiprobabilities \cite{Lostaglio2023,Wagner24,Buscemi2013,Rall2020}, based on a variety of techniques including weak measurements \cite{Resch2004,Mitchison2007,YungerHalpern2017}, or reconstruction of the KD characteristic function through interferometry \cite{Mazzola2013,YungerHalpern2017}.
In this work, we choose to employ the method presented in Ref.~\cite{Wagner24}.
This method is based on generalizing the standard swap test typically used to measure overlap between quantum states to a ``cycle test'' which measures arbitrary Bargmann invariants, including KD quasiprobabilities. 

\begin{figure}[t!]
    \begin{quantikz}[row sep={5mm, between origins},column sep=4mm]
    \lstick{$\ket{0}$} & \gate{H} & \ctrl{1} & \gate[style=dashed]{S} &\gate{H} & \meter{}
    \\[5mm]
    \lstick{$\ket{A}$} & & \gate[3]{\textsc{CYCLE}} &&
    \\
    \lstick{$\ket{\psi}$} & &  &&
    \\
    \lstick{$\ket{B}$} & &  &&
    \end{quantikz}
    \caption{\label{fig:CycleTest} Circuit to measure the real or imaginary part of the quasiprobability corresponding to the measurement outcomes $\dyad{A}$ and $\dyad{B}$ in the state $\ket{\psi}$. Which is measured is determined by the absence (real) or presence (imaginary) of the S gate. The controlled-cycle unitary performs a cyclic rotation of its inputs, taking $\ket{A}\ket{\psi}\ket{B} \to \ket{B}\ket{A}\ket{\psi}$.}
\end{figure}
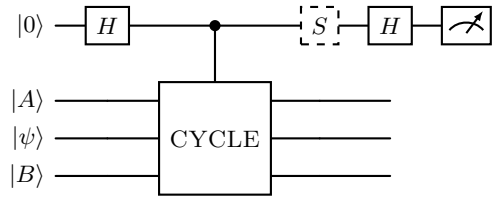

Specifically, given an ancilla qubit and three quantum registers initialized in states $\ket{A_i},\ket{B_j},\ket{\psi}$ the cycle test encodes the real or imaginary part of the quasiprobability $q_{ij}$ in the probability of finding the ancilla in its ground state $\ket{0}$,
\begin{subequations}
\begin{align}
    \RE{q_{ij}} &= \frac{1 + P_{0}(0)}{2}
    ,
    \\
    \IM{q_{ij}} &= \frac{1 + P_{\pi/2}(0)}{2}
    ,
\end{align}
\end{subequations}
where $P_{0,\pi/2}(0)$ correspond to circuits without or with an extra rotation on the ancilla before measurement as illustrated in Fig.~\ref{fig:CycleTest}.

%
%
\section{Model}
\label{sec:Model}

We will be using NISQ devices to simulate the dynamics of quantum models that do and do not support the emergence of classical behavior. Thus in choosing a model our goal is to have the ability to smoothly vary how non-classical the behavior of the system is, ranging from an effectively classical Darwinistic case to a case exhibiting strong quantum correlations.
Further, we require that the target system is simple enough to be simulated using current NISQ hardware.

\subsection{Hamiltonian}

For the purposes of ease of implementation, we consider a system-environment model consisting of a single system qubit (labeled $\sn{S}$) coupled to a collection of environment qubits (labeled $\sn{E}_i$) subject to a time-independent Hamiltonian.
For models of this type it is known precisely what constraints the Hamiltonian must satisfy to support Quantum Darwinism \cite{Doucet24,Duruisseau23}: the system Hamiltonian $\HHS$ specifies the pointer basis, the interaction Hamiltonian must factor as $\HHS \otimes \VV$, and the environment Hamiltonian must be a sum of local terms $\HHE = \sum_i {\HHE_j}$. Such models have been studied extensively in the context of Quantum Darwinism \cite{Zurek1982,Ollivier05,Zurek03,Duruisseau23,Ryan2022,Campbell2019}.

For the Hamiltonian, we choose
\begin{align}
    \HH 
    = \pqty{\frac{\Delta}{2} X^\sn{S} + \frac{\Omega}{2} Z^\sn{S}} \otimes \bbi^{\sn{E}}
    + X^{\sn{S}} \otimes \sum_{i=1}^{\NE} J_i X^{\sn{E}_i}
    .
    \label{eqn:Hamiltonian}
\end{align}
In the limit of a large environment $\NE\to\infty$ and at late times $t\to\infty$, the joint system-environment state approaches the singly-branching form and so classical objectivity emerges \cite{Touil24}, provided the coupling strengths $J_i$ are drawn from continuous random distributions and the transverse field $\Omega$ vanishes \cite{Doucet24,Duruisseau23}. 
If instead $\Omega > 0$, the Hamiltonian fails to support a pointer basis and so at late times the joint system-environment state will be some generic state with a more complex entanglement structure which cannot possibly support objectivity \cite{Touil24}. 

For experimental studies on NISQ devices, the asymptotic behavior with respect to $\NE$ or $t$ is not relevant and so we choose $\NE = 2$, which is as small as the environment can be while still supporting Darwinistic behavior with multiple observers \cite{Doucet25,Doucet24,Zurek1982}.
We further fix $\Delta = J = 1$, leaving only the transverse field strength $\Omega$ as a free parameter.
We always consider initial product states, as we are interested in correlations built by the Hamiltonian rather than those imposed by initial conditions.

\subsection{Simulations}

\begin{figure*}[ht]
    \centering
    \includegraphics[width=\textwidth]{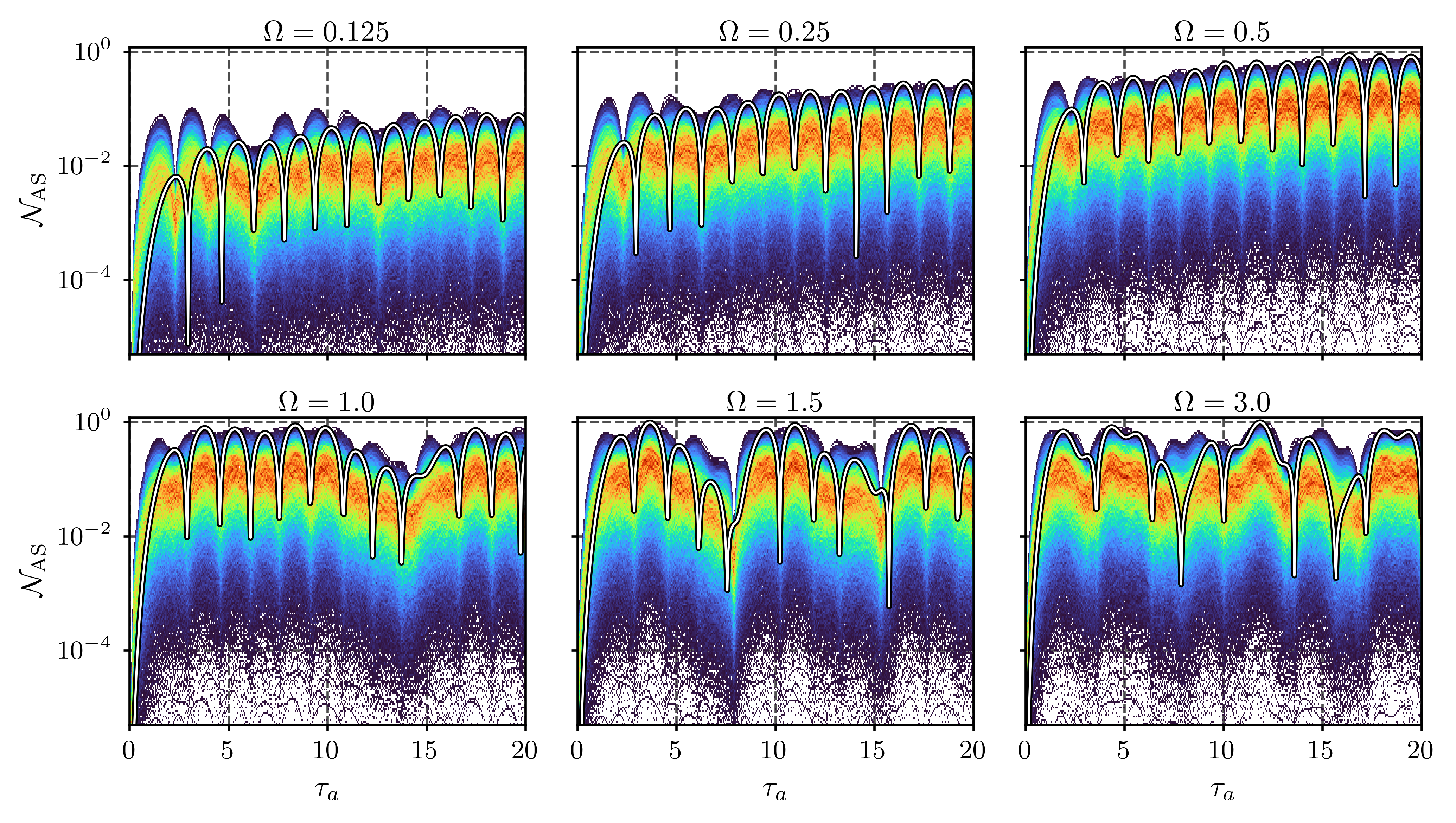}
    \caption{Heatmap of the non-classicality of the KDQ distribution over time for 5000 random settings, for a range of transverse field strengths $\Omega$. The white curve overlaid corresponds to the specific choice of settings used in our experiments, which approximately maximize the non-classicality at $\tau_a = 3.66$ when $\Omega = 1.5$.}
    \label{fig:MmtTraces}
\end{figure*}

Before testing this model using real hardware, we chose to explore its behavior through numerical simulation. 
The three-qubit Hamiltonian is sufficiently simple that it is trivial to compute the time evolution of any measurement operator, and hence it is straightforward to take an initial state and choice of measurement settings and produce the time-dependent quasiprobabilities.

We first check our assertion that for models with non-classical quasiprobabilities (and which therefore do not exhibit Quantum Darwinism), essentially all measurement choices should witness this fact.
To test this, we consider a pair of single-qubit measurements acting on qubits $\sn{E}_1$ and $\sn{E}_2$ taken to be at times $0\le\tau_a\le20$ and $0$, respectively. 
The measurement operators themselves as well as the initial three qubit product state were chosen at random using the respective Haar measures.
Figure~\ref{fig:MmtTraces} shows the distribution of non-classicality measures observed as the measurement time $\tau_a$ is swept, for a range of transverse field strengths $\Omega$.
In each panel, we observe an initial transient where the non-classicality builds from zero, with the initial growth rate faster when $\Omega$ is larger as expected. 
After this initial rise finite-size effects from the very small environment take over and we see some oscillations, with an infinite environment this should not be present.
In any case, we never observe the non-classicality measure to return to zero.

\begin{figure}[ht]
    \centering
    \includegraphics[width=0.5\textwidth]{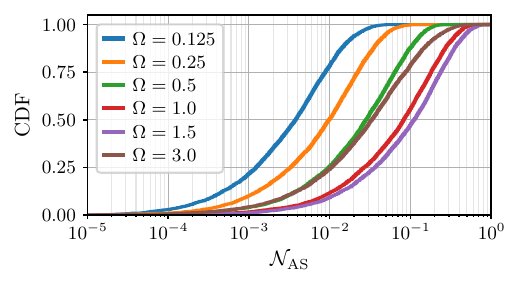}
    \caption{CDF of the distribution of non-classicality values for 5000 random settings with $\tau_a = 3.66$. Note that the probability of obtaining exactly zero non-classicality vanishes, indicating that for this simple system almost all measurement settings and initial conditions allow the failure of quantum Darwinism to be witnessed.}
    \label{fig:MmtHistogram}
\end{figure}

Figure~\ref{fig:MmtHistogram} shows the cumulative distribution functions of the non-classicality measure for various $\Omega$ with $\tau_a = 3.66$.
Notice that the probability of very small non-classicality measures $\mathcal{N}\lesssim10^{-5}$ goes to zero in all cases, indicating that all measurement settings show non-classical correlations on some level.
Conversely, similar tests performed with $\Omega = 0$ show $\mathcal{N} = 0$ always, as expected. 
This matches our expectation: if the model does not exhibit quantum Darwinism, collections of measurements with very low or vanishing non-classicality are exceedingly rare.

This demonstrates that while the existence of Quantum Darwinism requires $\ncl = 0$ for \emph{all} measurement choices and hence cannot be proven with a single measurement setting, for the simple model we consider here any generic choice can determine the presence or absence with very high probability.
We could therefore test this for Quantum Darwinism using a random measurement setting if we so desired, and that may be a reasonable course of action for systems which are not understood so thoroughly.
However, in this case we both fully understand our model system and wish to generate as much non-classicality as possible to maximize the ``signal-to-noise ratio'', providing the tested hardware platforms the best chance of distinguishing the cases accurately.
We therefore use the measurement setting corresponding to the white curve shown in Fig.~\ref{fig:MmtTraces}, which yields the largest $\ncl$ when $\Omega = 1.5$ and $\tau_a = 3.66$ while minimizing the complexity of the test circuit.

%
%
\section{Experiment}
\label{sec:Experiment}

In our case, since the projector $\Pi^A_j = \bbiS\otimes\pi^A_j\otimes\bbiB$ (and similarly for $\Pi^B_i$) are of rank 4, expanding the expression for $q_{ij}$ shows that it consists of 16 terms of the type which may be measured directly with the cycle test -- essentially, $q_{ij}$ comes from measurement of a more fine-grained set of quasiprobabilities which are then marginalized.
We may avoid the explicit need to marginalize through the introduction of additional ancilla qubits. 
For example, we augment the three qubit register in which we prepare $\ket{A}$ with two ancilla qubits. 
A Bell pair is prepared between the $\sn{S}$ qubit in the register and an ancilla, as well as between the $\sn{E}$ qubit and the other ancilla. 
Doing the same for the $\ket{B}$ register and tracing out the extra four ancilla qubits shows that the resulting circuit computes $q_{ij}$ directly, however scaled down by $1/16$ which reduces the accuracy to which it may be measured with a given shot count.
Increasing the number of shots by $16\times$ compensates for this, meaning that there is no overall performance benefit to this approach against measuring each term separately.
Indeed, tests with noisy simulators provided by IBM and IonQ show no substantial difference in accuracy between the two approaches, hence we choose the approach with extra ancilla for the conceptual clarity.

\begin{figure*}
    \begin{quantikz}[row sep={6mm, between origins},column sep=4mm]
    \lstick{$\ket{0}$} &[0.5cm] \gategroup[14,steps=5,style={inner
sep=6pt, dotted, rounded corners, fill=mplblu!20},background]{State Preparation} &&&&&[0.5cm] \gategroup[14,steps=1,style={inner
sep=6pt, dotted, rounded corners, fill=mplgrn!20},background]{Propagation} &[0.5cm] \gate{H}\gategroup[14,steps=9,style={inner
sep=6pt, dotted, rounded corners, fill=mplpur!20},background]{Cycle Test} & \ctrl{6} & \ctrl{7} & \ctrl{8} & \ctrl{1} & \ctrl{2} & \ctrl{3} & \gate[style=dashed]{S} &\gate{H} &[0.5cm] \meter{}
    \\[5mm]
    \lstick{$\ket{0}$} && \targ{} &&&& \gate[3][2cm]{~e^{i\HH \tau_a}}\gateinput{\scalebox{0.9}{$\sn{S}$}} &&&&& \swap{5} &&&&& \rstick[3]{$\ket{A}$}
    \\[-2mm]
    \lstick{$\ket{0}$} & \gate[style=dashed]{X} &&&&& \gateinput{\scalebox{0.9}{$\sn{E}_1$}} &&&&&& \swap{5} &&&&  
    \\[-2mm]
    \lstick{$\ket{0}$} &&& \targ{} &&& \gateinput{\scalebox{0.9}{$\sn{E}_2$}} &&&&&&& \swap{5} &&&
    \\
    \lstick{$\ket{0}$} & \gate{H} & \ctrl{-3} &&&&&&&&&&&&&&
    \\
    \lstick{$\ket{0}$} & \gate{H} && \ctrl{-2} &&&&&&&&&&&&&
    \\[3mm]
    \lstick{$\ket{0}$} &&&&&&&& \swap{3} &&& \swap{0} &&&&& \rstick[3]{$|\psi_0\rangle$}
    \\[-2mm]
    \lstick{$\ket{0}$} &&&&&&&&& \swap{3} &&& \swap{0} &&&&
    \\[-2mm]
    \lstick{$\ket{0}$} &&&&&&&&&& \swap{3} &&& \swap{0} &&&
    \\[3mm]
    \lstick{$\ket{0}$} && \targ{} &&&&&& \swap{0} &&&&&&&& \rstick[3]{$\ket{B}$}
    \\[-2mm]
    \lstick{$\ket{0}$} &&& \targ{} &&&&&& \swap{0} &&&&&&&
    \\[-2mm]
    \lstick{$\ket{0}$} & \gate[style=dashed]{X} &&& \gate{H} & \gate{S} &&&&& \swap{0} &&&&&&
    \\
    \lstick{$\ket{0}$} & \gate{H} & \ctrl{-3} &&&&&&&&&&&&&&
    \\
    \lstick{$\ket{0}$} & \gate{H} && \ctrl{-3} &&&&&&&&&&&&&
    \end{quantikz}
    \caption{\label{fig:FullCircuit} Circuit diagram showing the complete experimental protocol. The first section of the circuit prepares the desired states in the three-qubit registers representing $\ket{A}$, $\ket{B}$, and $\ket{\psi}$. Then, the state $\ket{A}$ is evolved with the Trotterized propagator $U^\dagger$ to reach the specified measurement time $\tau_a$. Finally, the cycle test of \cite{Wagner24} is applied to measure the quasiprobability $q = \TR{BA(\tau_a)\dyad{\psi}}$.}
\end{figure*}
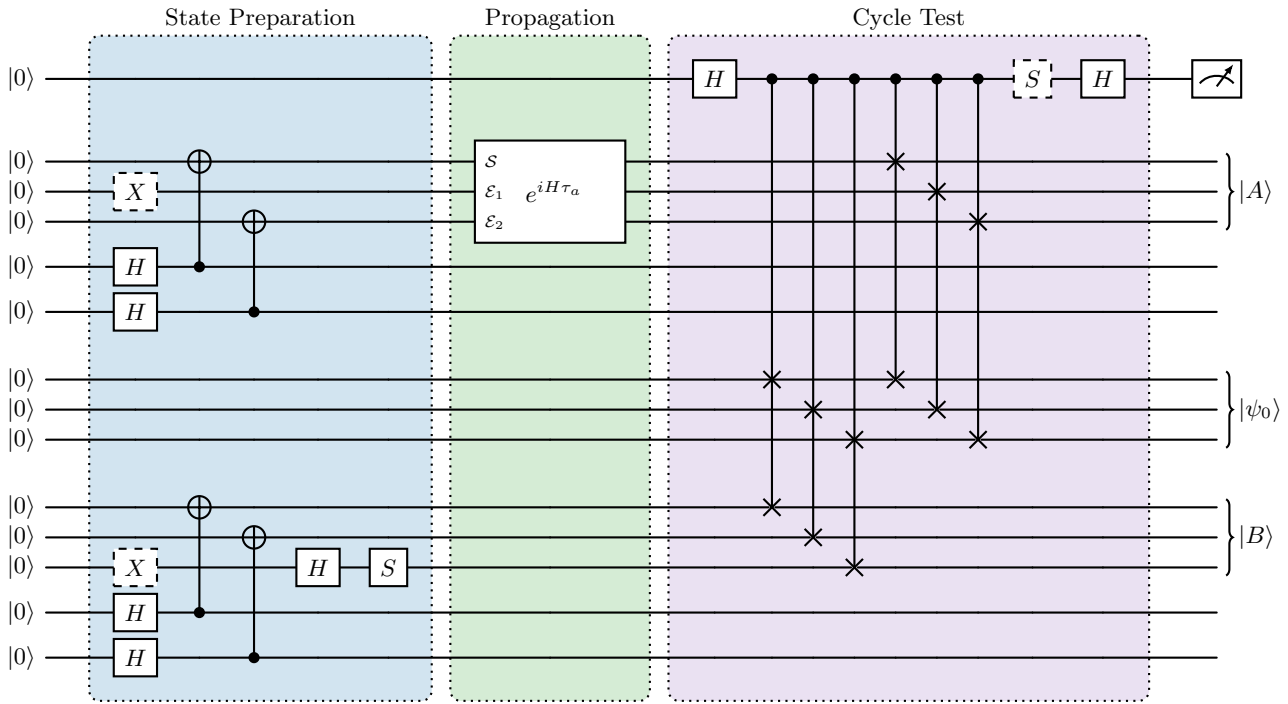

The circuit we employ is presented in Fig.~\ref{fig:FullCircuit}, broken down into three stages. 
In the first stage, we prepare the states corresponding to the projectors measured by Alice and Bob in two registers, corresponding to their respective $X$ or $Y$ bases. 
Which projector is prepared is selected by the inclusion or exclusion of the dashed $X$ gate. 
The initial state $\ket{\psi_0} = \ket{000}$ requires no action to prepare.
Subsequently, the register representing Alice's projector is evolved with the propagator for $\HH$ for a time $\tau_a$, decomposed with five Trotter steps.
From simulations we find that this number of steps is required to allow accurate measurements of the quasiprobabilities.
The simplicity of the Hamiltonian ensures that the circuit remains simple to implement, nonetheless.
Finally, the cycle test is applied to the three registers and the quasiprobability is encoded into $P(0)$ of the first ancilla qubit. 
The real part is returned if the dashed $S$ gate is absent, the imaginary part if it is present.

It should be noted that this technique for measurements of quasiprobabilities in Darwinistic systems scales poorly. 
For example, if we considered a four qubit model, the factor by which $q_{ij}$ is scaled (equivalently, the number of individual terms to measure) would be $64$. 
Every qubit in the model which is not ``measured'' by either Alice or Bob adds a factor of $4$, either to the shot count or to the circuit count.
In the future when larger models are considered, it will be necessary to consider other approaches to measuring quasiprobabilities or otherwise testing for non-classicality.
However, as we will demonstrate, the overhead is not unreasonable for small models. 
Additionally, it is clear that other factors limit what is achievable with current NISQ devices and so this approach provides a useful benchmark even when only considering very small systems.

\subsection{Experimental Procedure}

In this work, we ran all variations of the circuit of Fig.~\ref{fig:ExperimentResults} on both superconducting and trapped-ion platforms, orchestrated with the {\tt qiskit} library. 
Specifically, we made use of a 133 qubit IBM Heron superconducting quantum computer with a heavy-hexagon topology, the {\tt ibm\_torino} device, as well as the IonQ {\tt Aria-1} device, a 25 qubit trapped-ion machine with all-to-all connectivity.
The physical parameters of the two devices are presented in Appendix~\ref{app:Characterization} for reference.

On both devices, we ran 10000 shots of all eight configurations of the circuit of Fig.~\ref{fig:ExperimentResults} to measure the real and imaginary parts of all four quasiprobabilities $q_{ij}$. 
This procedure was repeated three times (24 circuits total), once with $\tau_a = 3.66$, once with $\tau_a = 2.21$, and once with the propagation step removed entirely corresponding to $\tau_a = 0$. 

Additionally, both companies provide noisy simulators which aim to allow testing and calibration to be performed before running on real hardware. 
Before running on real hardware we ran the same 24 circuits on the noisy simulators to validate that our approach was likely to work with this particular choice of Hamiltonian parameters, measurement settings, and shot counts.

\subsection{Results}

\begin{figure}[ht]
    \centering
    \includegraphics[width=\columnwidth]{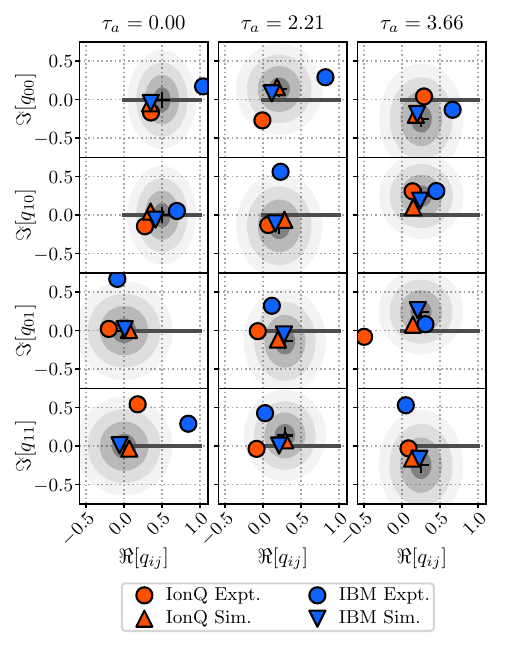}
    \caption{Scatterplot of the complex quasiprobabilities measured using the circuit of Fig.~\ref{fig:FullCircuit} for inter-measurement times of (top) $\tau_a = 0$, (middle) $\tau_a = 2.21$, and (bottom) $\tau_a = 3.66$. Results are shown for experiments using the IonQ {\tt Aria-1} and IBM {\tt Torino} quantum computers with $2\times10^5$ shots per quasiprobability, as well as noisy simulations of the same machines with $2\times10^6$ shots per quasiprobability. The numerically exact quasiprobabilities are depicted using crosses, with shaded ellipses denoting the expected $1\sigma,2\sigma,3\sigma,4\sigma$ regions assuming only statistical uncertainty inherent to estimating $q_{ij}$ with finitely many samples.}
    \label{fig:ExperimentResults}
\end{figure}

Our results are summarized in Fig.~\ref{fig:ExperimentResults}. We plot the quasiprobabilities measured on each device or simulator for each set of parameters on the complex plane, compared to the exact quasiprobability computed through classical simulation with the definition of the KD distribution.
The results of the noisy simulations of each device are fairly close to the theoretical value for the quasiprobabilities, indicating that the circuits we run should be able to provide accurate measurements with the shot counts employed. 

Unfortunately, the results obtained with the two quantum computers show that these noisy simulations are overly optimistic in their predictions.
In Tab.~\ref{tab:RMSE} we quantify this by computing the root mean square error of each set of four measured quasiprobabilities, which is much higher for the devices compared to the simulations.

\begin{table}
    \centering
    \begin{tabularx}{0.90\columnwidth}{@{}r*3{>{\centering\arraybackslash}X}@{}}
        \toprule
        & $\tau_a = 0$ & $\tau_a = 2.21$ & $\tau_a = 3.66$ \\
        \midrule
        \midrule
        IonQ Sim. & 0.294 & 0.056 & 0.159 \\
        Expt. & 0.491 & 0.364 & 0.461 \\
        \midrule
        IBM Sim. & 0.333 & 0.127 & 0.070 \\
        Expt. & 0.777 & 0.563 & 0.476 \\
        \bottomrule
    \end{tabularx}
    \caption{Root mean square error in the quasiprobabilities measured with each device and each simulator.}
    \label{tab:RMSE}
\end{table}

Finally, we tabulate the non-classicality measure of Eq.~\eqref{eqn:NclAS} as determined for each experiment in Tab.~\ref{tab:Nonclassicality} alongside the exact value. 
This quantity is quite sensitive to errors in the quasiprobabilities, so based on the accuracy we were able to attain in measuring those quasiprobabilities it is not surprising that the values for $\mathcal{N}_{\rm AS}$ we find do not match the exact result.
With the noisy simulator the results are somewhat better -- especially when $\tau_a = 0$ and the time evolution step is removed from the circuit -- however the values returned for other $\tau_a$ are still quite inaccurate.

That said, it is clear that derived quantities such as $\mathcal{N}_{\rm AS}$ are essentially out of reach so long as individual quasiprobabilities cannot be measured accurately.
Additionally, we note that on both the IBM and IonQ devices additional tests with increased shot counts do not show improvements in the results.
This indicates that the inaccuracy of our results is not just an artifact of sampling imprecision, but that the noise level of these devices is a significant limiting factor on the achievable accuracy.

\begin{table}
    \centering
    \begin{tabularx}{0.90\columnwidth}{@{}r*3{>{\centering\arraybackslash}X}@{}}
        \toprule
        & $\tau_a = 0$ & $\tau_a = 2.21$ & $\tau_a = 3.66$ \\
        \midrule
        \midrule
        Theory & 0.000 & 0.554 & 0.988 \\
        \midrule
        IonQ Sim. & -0.041 & 0.364 & 0.135 \\
        Expt. & 0.888 & -0.328 & 0.472 \\
        \midrule
        IBM Sim. & -0.033 & 0.007 & 0.675 \\
        Expt. & 2.869 & 1.800 & 1.531 \\
        \bottomrule
    \end{tabularx}
    \caption{Non-classicality measure $\mathcal{N}_{AS}$ for each parameter set as computed using the IBM {\tt Torino} and IonQ {\tt Aria-1} devices, compared with noisy simulations and exact results.}
    \label{tab:Nonclassicality}
\end{table}

%
%
\section{Concluding remarks}
\label{sec:Conclusions}
In conclusion, we have leveraged previous works connecting Quantum Darwinism, emergent classicality, and properties of KD quasiprobability distributions to build a method to study the quantum-to-classical transition in a simple qubit model experimentally using NISQ hardware. 
At present, this problem serves as a practical benchmark for quantum computers -- how non-classical must a simulated system be before it can be certified as such?
And how precisely can the degree to which a simulated system is non-classical be quantified in practice?
Our results using the noisy simulators provided by IBM and IonQ suggest that it should be possible to achieve good resolution using current hardware.
Looking forward, this points to a direction for understanding Quantum Darwinism and the fuzzy boundary between the quantum and classical worlds experimentally for a variety of systems.
This can be a useful tool for deepening our understanding, especially when studying the dynamics of increasingly larger systems as the available devices grow in size and capability.

On the other hand, our experimental results do not rise to the level of what is predicted by the simulators. 
Clearly this points to there being room for improvement in the noise models employed by the simulators, but this stark difference also suggests that this problem may serve as a useful benchmark for NISQ devices. One important advantage to the problem examined in this work as a benchmark is that it is not tied to any specific quantum circuit or gate set, and may even be adaptable to platforms implementing other paradigms of quantum information processing.
In this sense it is a somewhat ``high-level'' benchmark, compared to for example randomized benchmarking \cite{Helsen2022,Proctor2017,Knill2008} used to characterize error rates in quantum gates.

In any case, by asking detailed questions about how precisely a quantum computer can quantify non-classicality we can make statements about these devices applicability to various simulation tasks, and potentially comment on the algorithmic competition between noise and quantum advantage.

\acknowledgments{E.D. acknowledges U.S. NSF under Grant No. OSI-2328774.}

\appendix

\section{Device Characterization}
\label{app:Characterization}

In Tab.~\ref{tab:DeviceParameters} we present the median decoherence rates, gate times, and gate error rates for the IBM {\tt Torino} and IonQ {\tt Aria-1} devices. Note that these values reported by IBM and IonQ are based on regular calibrations and may change day by day.

\begin{table}[b!]
    \centering
    \begin{tabularx}{0.90\columnwidth}{@{}r*2{>{\centering\arraybackslash}X}@{}}
        \toprule
        & \textbf{IBM \texttt{Torino}} & \textbf{IonQ {\tt Aria-1}}\\
        \midrule
        \midrule
        Topology & Heavy hexagon & All-to-all \\
        \midrule
        $T_1$ Time & \SI{183.29}{\micro\second} & \SI{100}{\second} \\
        $T_2$ Time & \SI{141.73}{\micro\second} & \SI{1}{\second} \\
        1Q Gate Time & \SI{32}{\nano\second} & \SI{135}{\micro\second} \\
        2Q Gate Time & \SI{68}{\nano\second} & \SI{600}{\micro\second} \\
        Readout Time & \SI{1560}{\nano\second} & \SI{300}{\micro\second} \\
        \midrule
        1Q Gate Error & 0.037\% & 0.010\% \\
        2Q Gate Error & 0.286\% & 1.240\% \\
        Readout Error & 2.539\% & -- \\
        SPAM Error & -- & 0.370\% \\
        \bottomrule
    \end{tabularx}
    \caption{Performance characterizations of the 133 qubit IBM {\tt Torino} device \cite{IBMCharacterization} and 25 qubit IonQ {\tt Aria-1} device \cite{IonQCharacterization}. All parameters reported are median values across all qubits.}
    \label{tab:DeviceParameters}
\end{table}

\bibliography{references}
\end{document}